\documentclass[aps,prl,reprint,superscriptaddress]{revtex4-1}
\usepackage{graphicx,xcolor}
\usepackage{amsmath}
\usepackage{braket}
\usepackage{url}
\usepackage{filecontents}

\begin{document}

\title{Evolution of antiferromagnetic domains in the all-in-all-out ordered pyrochlore Nd$_2$Zr$_2$O$_7$}

\author{L. Opherden}
\email[]{l.opherden@hzdr.de}
\author{J. Hornung}
\affiliation{Hochfeld-Magnetlabor Dresden (HLD-EMFL), Helmholtz-Zentrum Dresden-Rossendorf,	01314 Dresden, Germany}
\affiliation{Institut f\"{u}r Festk\"orperphysik, TU Dresden, 01062 Dresden, Germany}
\author{T. Herrmannsd\"orfer}
\affiliation{Hochfeld-Magnetlabor Dresden (HLD-EMFL), Helmholtz-Zentrum Dresden-Rossendorf,	01314 Dresden, Germany}
\author{J. Xu}
\affiliation{\mbox{Helmholtz-Zentrum Berlin f\"{u}r Materialien und Energie GmbH, Hahn-Meitner Platz 1, 14109 Berlin, Germany}}
\affiliation{\mbox{Institut f\"{u}r Festk\"{o}rperphysik, Technische Universit\"{a}t Berlin, Hardenbergstra$\beta$e 36,10623 Berlin, Germany}}
\author{A.~T.~M.~N. Islam}
\affiliation{\mbox{Helmholtz-Zentrum Berlin f\"{u}r Materialien und Energie GmbH, Hahn-Meitner Platz 1, 14109 Berlin, Germany}}
\author{B. Lake}
\affiliation{\mbox{Helmholtz-Zentrum Berlin f\"{u}r Materialien und Energie GmbH, Hahn-Meitner Platz 1, 14109 Berlin, Germany}}
\affiliation{\mbox{Institut f\"{u}r Festk\"{o}rperphysik, Technische Universit\"{a}t Berlin, Hardenbergstra$\beta$e 36,10623 Berlin, Germany}}
\author{J. Wosnitza}
\affiliation{Hochfeld-Magnetlabor Dresden (HLD-EMFL), Helmholtz-Zentrum Dresden-Rossendorf,	01314 Dresden, Germany}
\affiliation{Institut f\"{u}r Festk\"orperphysik, TU Dresden, 01062 Dresden, Germany}

\date{\today}

\begin{abstract}
We report the observation of magnetic domains in the exotic, antiferromagnetically ordered all-in-all-out state of Nd$_2$Zr$_2$O$_7$, induced by spin canting.
The all-in-all-out state can be realized by Ising-like spins on a pyrochlore lattice and is established in Nd$_2$Zr$_2$O$_7$ below 0.31 K for external magnetic fields up to 0.14 T. Two different spin arrangements can fulfill this configuration which leads to the possibility of magnetic domains. The all-in-all-out domain structure can be controlled by an external magnetic field applied parallel to the [111] direction. This is a result of different spin canting mechanism for the two all-in-all-out configurations for such a direction of the magnetic field. The change of the domain structure is observed through a hysteresis in the magnetic susceptibility.
No hysteresis occurs, however, in case the external magnetic field is applied along [100]. 
\end{abstract}


\maketitle


Cubic pyrochlore oxides, $R_2T_2$O$_7$ ($R$ = rare-earth element, $T$ = transition metal), have attracted strong interest since the discovery of spin ice in Ho$_2$Ti$_2$O$_7$  and Dy$_2$Ti$_2$O$_7$ \cite{Harris_1997, Bramwell_2001, Gardner_2010, Siddharthan_1999, Snyder_2001} and the excitation of magnetic monopoles \cite{Castelnovo_2008, Castelnovo_2011, Khomskii_2012} even though their low temperature properties are known since more than fifty years \cite{Blote_1969-2}. In these systems, the $R$ and $T$ ions are each located on a sublattice of corner-sharing tetrahedra.
In pyrochlore compounds the strong crystal electrical field (CEF) mostly splits up the 2$J$+1 multiplet of the $R$ ion by several hundred kelvin which can result in an effective spin-half ground state with a strong magnetic anisotropy \cite{Gardner_2010}. The spins of the magnetic ions are highly anisotropic, and can behave as Ising spins forced to point mostly along the local $\langle$111$\rangle$ direction, i.e., the four corner-to-center directions of each tetrahedron, or as $XY$ spins as in Er$_2$Ti$_2$O$_7$ \cite{Dasgupta_2006}.

The interplay between exchange (nearest-neighbor (nn) and next-nearest-neighbor) interaction and dipole-dipole (and higher multipoles) interaction leads to a variety of ground states, such as ferromagnetically \cite{Yasui_2001, Zhou_2008-1} or antiferromagnetically \cite{Champion_2001} ordered states, spin ice \cite{Harris_1997, Bramwell_2001, Kadowaki_2002, Anand_2016-1} or the absence of long-range order down to the lowest temperatures as found in spin liquids \cite{Gingras_2000, Sibille_2015}.

Besides the intensively studied spin-ice regime, the all-in-all-out state just recently became a highly discussed topic due to its proximity to the spin-ice regime.
The all-in-all-out configuration is an antiferromagnetically ordered state where all spins of a given tetrahedron are pointing either inward towards its center or outward. This order can be realized by two spin arrangements, either All-In-All-Out (AIAO) or All-Out-All-In (AOAI), which are interchangeable by time-reversal transformation \cite{Arima_2012} (see Fig. \ref{fig:AIAO-Canting_visualization}). The existence of these two spin arrangements allows for the formation of domains within this ordered ground state as it was shown experimentally \cite{Ma_2015, Tardif_2015}. If the degeneracy of this state can be lifted by an external magnetic field, the order becomes ferrimagnetic \cite{Arima_2012} and the domain structure can be changed.

$R_2T_2$O$_7$ systems which exhibit such a ground-state configuration are for instance Nd$_2$Zr$_2$O$_7$ \cite{Xu_2015-1, Lhotel_2015}, Nd$_2$Hf$_2$O$_7$ \cite{Anand_2015}, Nd$_2$Sn$_2$O$_7$ \cite{Bertin_2015}, and Nd$_2$Ir$_2$O$_7$. In Nd$_2$Ir$_2$O$_7$, the transition to the all-in-all-out phase is accompanied by a metal-insulator transition, the occurrence of a huge magnetoresistance and the stabilization of a Weyl semimetallic state \cite{Tian_2016}.

In contrast to Nd$_2$Ir$_2$O$_7$, where both, the Nd$^{3+}$ and the Ir$^{4+}$ are magnetic, in Nd$_2$Zr$_2$O$_7$, only the sublattice of the Nd$^{3+}$ ion forms the all-in-all-out order. 
In addition, we determined the ordering temperature of Nd$_2$Zr$_2$O$_7$ ($T_N$ = 0.31 K) to be two magnitudes smaller than in Nd$_2$Ir$_2$O$_7$ ($T_N$ $\sim$ 32 K \cite{Ma_2015}). Because of this difference in the energy scale, Nd$_2$Zr$_2$O$_7$ is an adequate model system to study the magnetic properties of the exotic all-in-all-out ground state using small magnetic fields.
Furthermore, this compound recently gained interest due to results from neutron scattering experiments. They showed evidence for the coexistence of all-in-all-out ordering and spin-ice behavior for which the concept of fragmentation of the magnetic moment was proposed \cite{Petit_2016-1}.

Here, we report that the all-in-all-out domain structure of Nd$_2$Zr$_2$O$_7$ can be controlled by an external magnetic field. We probe the spin and domain dynamics in the all-in-all-out ground state configuration by means of dynamic susceptibility and static magnetization measurements where we find that the all-in-all-out order is established below $T_N$ = 0.31 K and is stable for external magnetic fields up to $\mathrm{\mu}_0 H_{dc}$~=~0.14 T.

Static (DC) magnetization was measured using a DC SQUID magnetometer equipped with a dilution refrigerator moving through a 2$^{\mathrm{nd}}$ order gradiometer. Dynamic (AC) susceptibility was measured with a pair of compensated coils using frequencies between 20~Hz and 25~kHz and field amplitudes, $H_{ac}$, between 75 nT and 10 $\mathrm{\mu}$T. The AC-field direction was aligned parallel to the DC field.
If not denoted differently, the AC susceptibility was measured at $f$ = 2500 Hz and $\mathrm{\mu}_0 H_{ac}$~=~2.8~$\mathrm{\mu}$T. The measurements were performed on two Nd$_2$Zr$_2$O$_7$ single crystals with different crystal orientations, both having a weight of approximately 8~mg.  The single crystals were grown by the floating-zone technique using a high-temperature optical image furnace.


\begin{figure}[t]
	\centering
	\includegraphics[width=\columnwidth]{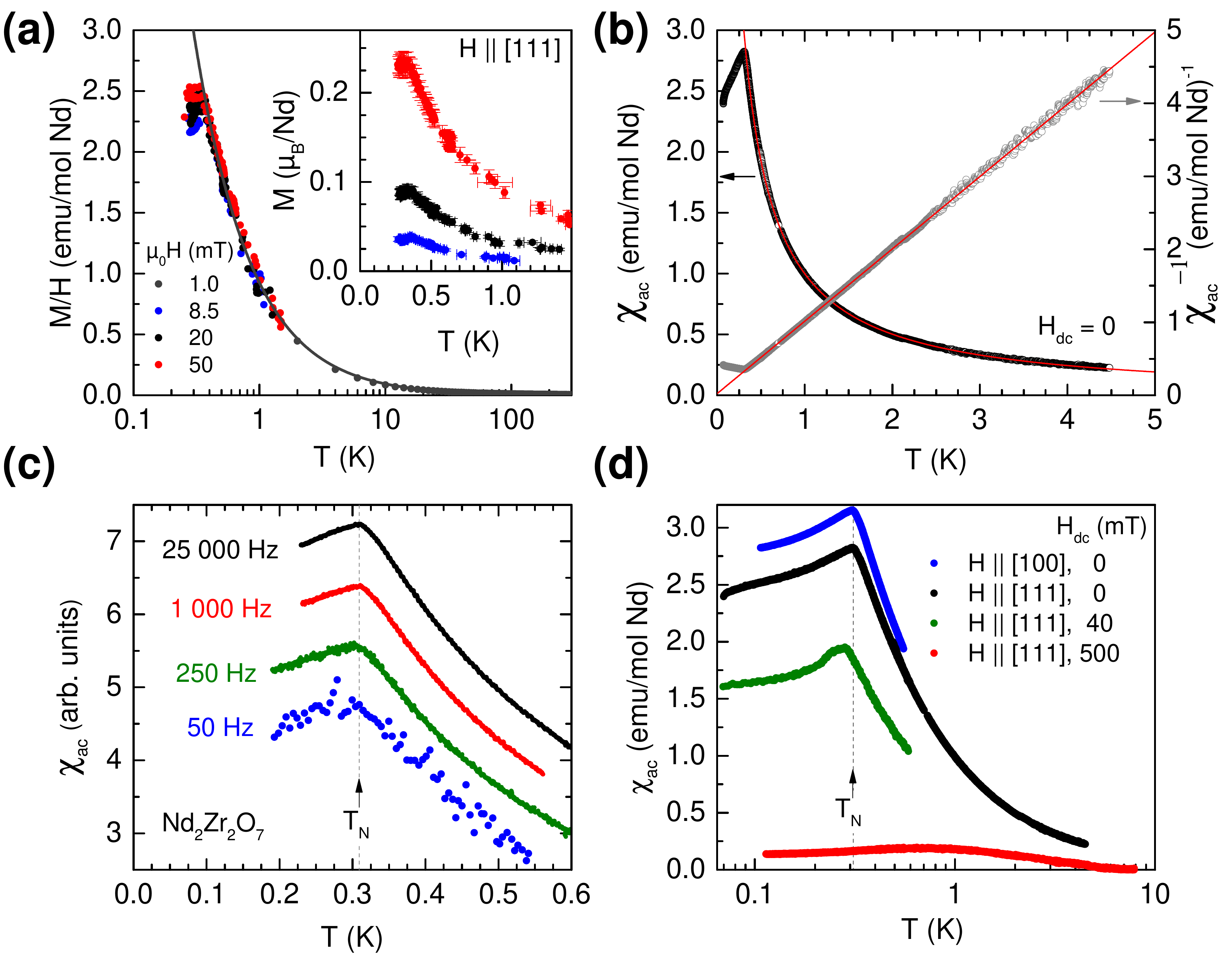}
	\caption{\textbf{(a)} Static magnetic susceptibility versus temperature for Nd$_2$Zr$_2$O$_7$, H $\|$ [111], for small magnetic fields. A Curie-Weiss fit with $\theta = - 0.01$ K describes the low-temperature behavior above $T_N$ (fit range $2$ K $< T < 50$ K). Inset: Magnetization of the same data.\textbf{(b)} AC susceptibility and inverse AC susceptibility versus Temperature for $H_{ac} \|$  [111] as well as the corresponding Curie-Weiss fit (fit range $0.4$ K $< T < 4.5$ K). \textbf{(c)} AC susceptibility versus temperature for different frequencies. \textbf{(d)} Comparison of the AC susceptibility versus temperature for different orientations of the crystal with respect to the AC field as well as for an external DC field of 0.04 T and 0.5 T. For $\mu_0 H_\mathrm{dc}$ = 0.04 T, data taken at $f$ = 160 Hz was used.} 
	\label{fig:Tempsweep_SQUID+ACS}
\end{figure}
Both, the temperature dependence of the static susceptibility, $\chi_\mathrm{stat} \approx M / H$ and the amplitude of the AC susceptibility $\chi_\mathrm{ac} = \sqrt{\chi^{\prime 2} + \chi^{\prime\prime 2}}$ follow the Curie-Weiss law $\chi = \frac{C}{T - \theta}$ with a paramagnetic Curie temperature around zero, $\theta~=~-0.01 \pm 0.03$~K [Fig. \ref{fig:Tempsweep_SQUID+ACS} (a), (b)]. The small value of the paramagnetic Curie temperature indicates that $J_\mathrm{nn}$ and $D_\mathrm{nn}$ are balanced, and that Nd$_2$Zr$_2$O$_7$ lies on the border between all-in-all-out ordering and the spin-ice regime.

At $T_N$~=~0.31~K, Nd$_2$Zr$_2$O$_7$ shows a maximum in the zero-field AC susceptibility for $H_{ac} \|$~[111] as well as for $H_{ac} \|$~[100] [Fig. \ref{fig:Tempsweep_SQUID+ACS} (d)].
Both peaks, in $\chi^{\prime}$ and in $\chi^{\prime\prime}$, appear at the same temperature (not shown).

A straightforward approach to describe the magnetic interactions on the pyrochlore lattice is the dipolar spin-ice model, where the Hamiltonian considers only the dominant nn exchange, $J_\mathrm{nn}$, and the dipole energy $D_\mathrm{nn}$,
\begin{multline}
H_\mathrm{DSI} = -J_\mathrm{nn} \cdot 3 \sum_{\langle ij\rangle} \mathbf{S}_i^{z_i} \cdot \mathbf{S}_j^{z_j} \\
+ D_\mathrm{nn} \frac{3 r^3_\mathrm{nn}}{5} \sum_{j<i} \frac{ \mathbf{S}_i^{z_i} \cdot \mathbf{S}_j^{z_j}}{|\mathbf{r}_{ij}|^3} - \frac{3(\mathbf{S}_i^{z_i}\cdot\mathbf{r}_{ij})(\mathbf{S}_j^{z_j}\cdot\mathbf{r}_{ij})}{| \mathbf{r}_{ij}|^5} .
\end{multline}
In this model, a spin ice or an all-in-all-out ground-state configuration is favored depending on the ratio of $J_\mathrm{nn}/D_\mathrm{nn}$ \cite{denHertog_2000, Melko_2004, Brooks-Bartlett_2014}.
The latter is only achieved if $  J_\mathrm{nn} / D_\mathrm{nn} <  -0.91 $ \cite{denHertog_2000}, which requires an antiferromagnetic exchange interaction.
The disadvantage of this model is, that the spins are treated as ideal Ising spins which is only approximately the case for Nd spins in Nd$_2$Zr$_2$O$_7$.
Instead, Nd$^{3+}$ is, besides Dy$^{3+}$, the only element of the lanthanide series whose $J$ value and CEF parameter $B^0_2$ allows the Kramers ground-state doublet to be a dipolar-octupolar doublet \cite{Huang_2014}.
In this case, two components of the effective pseudospin operator $\vec{\tau}$ behave like a dipole under space-group transformation (including the dominant component, which couples to $J_\mathrm{z}$ that points along the local $\langle$111$\rangle$ direction), whereas the third component behaves like an octupolar tensor. Y.-P. Huang \textit{et al. } introduced the $XYZ$ model for the dipolar-octupolar doublet in the localized limit \cite{Huang_2014}.
Depending on the ratio between the components of $\vec{J}$, quantum spin ice, antiferro-octupolar ordering or all-in-all-out ordering can be established at lowest temperatures.

Evidence, that in the case of Nd$_2$Zr$_2$O$_7$ the all-in-all-out ground-state configuration is realized, was recently provided by neutron diffraction \cite{Petit_2016-1, Lhotel_2015, Xu_2015-1} and mean-field calculations \cite{Lhotel_2015}.
Therefore, we can conclude that the peak of the AC susceptibility at $T_N$ is the signature of the phase transition to the all-in-all-out ground state. Further evidence is gained from the observation that the maximum in $\chi_\mathrm{ac}$($T$) is shifted to lower temperatures by applying a small external magnetic field and completely suppressed using a moderate external magnetic field, such as 0.5~T along the [111] direction [Fig. \ref{fig:Tempsweep_SQUID+ACS} (d)]. 
Compared to spin-ice systems, where typically a maximum of the AC susceptibility appears at a certain temperature which is strongly frequency dependent \cite{Sibille_2016-1, Matsuhira_2000, Snyder_2004}, Nd$_2$Zr$_2$O$_7$ does not show any frequency dependence of $\chi_\mathrm{ac}$($T$) for frequencies of 50 Hz up to 25 kHz [Fig. \ref{fig:Tempsweep_SQUID+ACS} (c) ] consistent with a previous work, reporting no dependence between 0.11 and 570 Hz \cite{Lhotel_2015}. Furthermore, no additional features can be seen in the AC susceptibility data at temperatures up to 4 K. Therefore, we have to underline that we do not find evidence for the spin-ice phase in coexistence with the all-in-all-out phase, which was reported to exist in Nd$_2$Zr$_2$O$_7$ in consequence of observing a pinch-point pattern which persists up to 0.6 K \cite{Petit_2016-1}. It is puzzling why a Coulomb phase, emerging from the fragmentation of the magnetic moment, would not be observable by means of dynamic-susceptibility measurements. Furthermore, the theory of magnetic-moment fragmentation predicts that the characteristic cusp of the antiferromagnetic phase transition would be masked and the AC susceptibility would appear to be featureless which is contradiction to the clear visible cusp-like feature appearing at $T_N$ [Fig. \ref{fig:Tempsweep_SQUID+ACS} (a) - (d)].

\begin{figure}[t]
	\centering
	\includegraphics[width=\columnwidth]{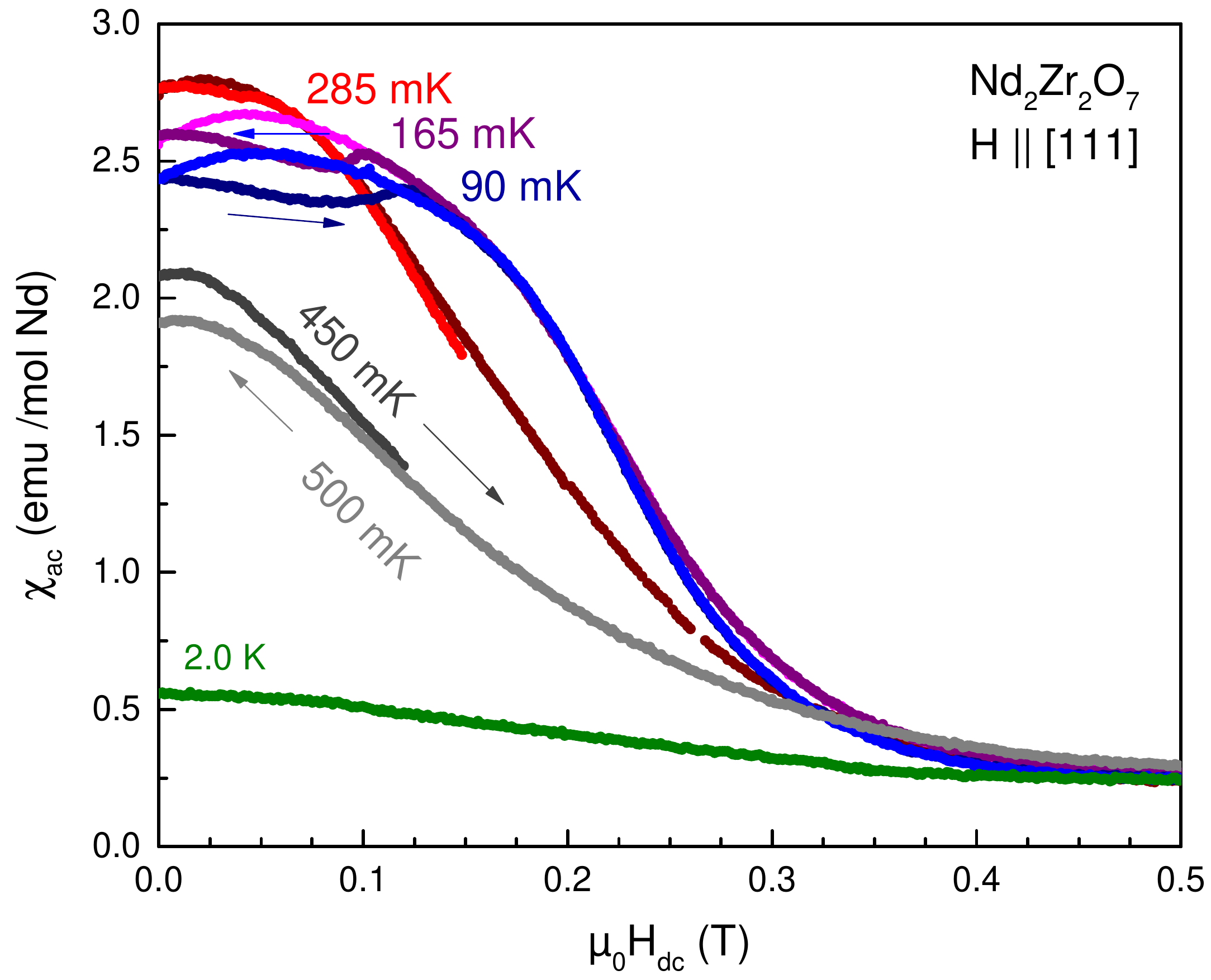}
	\caption{AC susceptibility versus external field applied along the [111] direction. Measurements were performed at various temperatures.}
	\label{fig:X_vs_B_Hpar111_all}
\end{figure}
Below $T_N$, a hysteresis occurs in the magnetic-field dependence of the AC susceptibility, for fields aligned along [111] (Fig. \ref{fig:X_vs_B_Hpar111_all}). After zero-field cooling the susceptibility first decreases with increasing DC field, running through a shallow and broad minimum.
At a certain field, $H^\ast$, a maximum appears after which $\chi_\mathrm{ac}$($H$) decreases again for higher DC fields. This maximum is the signature of a spin-flop transition where the system enters the 3-in-1-out or 2-in-2-out configuration depending on the field direction \cite{Tian_2016, Lhotel_2015}.

Reducing $H_{dc}$ from the polarized state ($H_{dc} > H^\ast$), however, the susceptibility shows a broad maximum instead of the minimum, before $\chi_\mathrm{ac}$($H$) reaches its zero-field value at $H$ = 0. 
A qualitative similar behavior was reported for the derivative of the static magnetization $\left(\mathrm{d}M/\mathrm{d}H\right)$ of the all-in-all-out-ordered system Nd$_2$Ir$_2$O$_7$  but not discussed further \cite{Tian_2016}. In contrast to Nd$_2$Ir$_2$O$_7$, the $T^{4+}$-ions (Zr$^{4+}$) in Nd$_2$Zr$_2$O$_7$ possess no magnetic moment. The hysteresis is therefore a direct result of the all-in-all-out state and not attributed to the interplay of the two all-in-all-out ordered sublattices in Nd$_2$Ir$_2$O$_7$.

\begin{figure}[t]
	\centering
	\includegraphics[width=0.85\columnwidth]{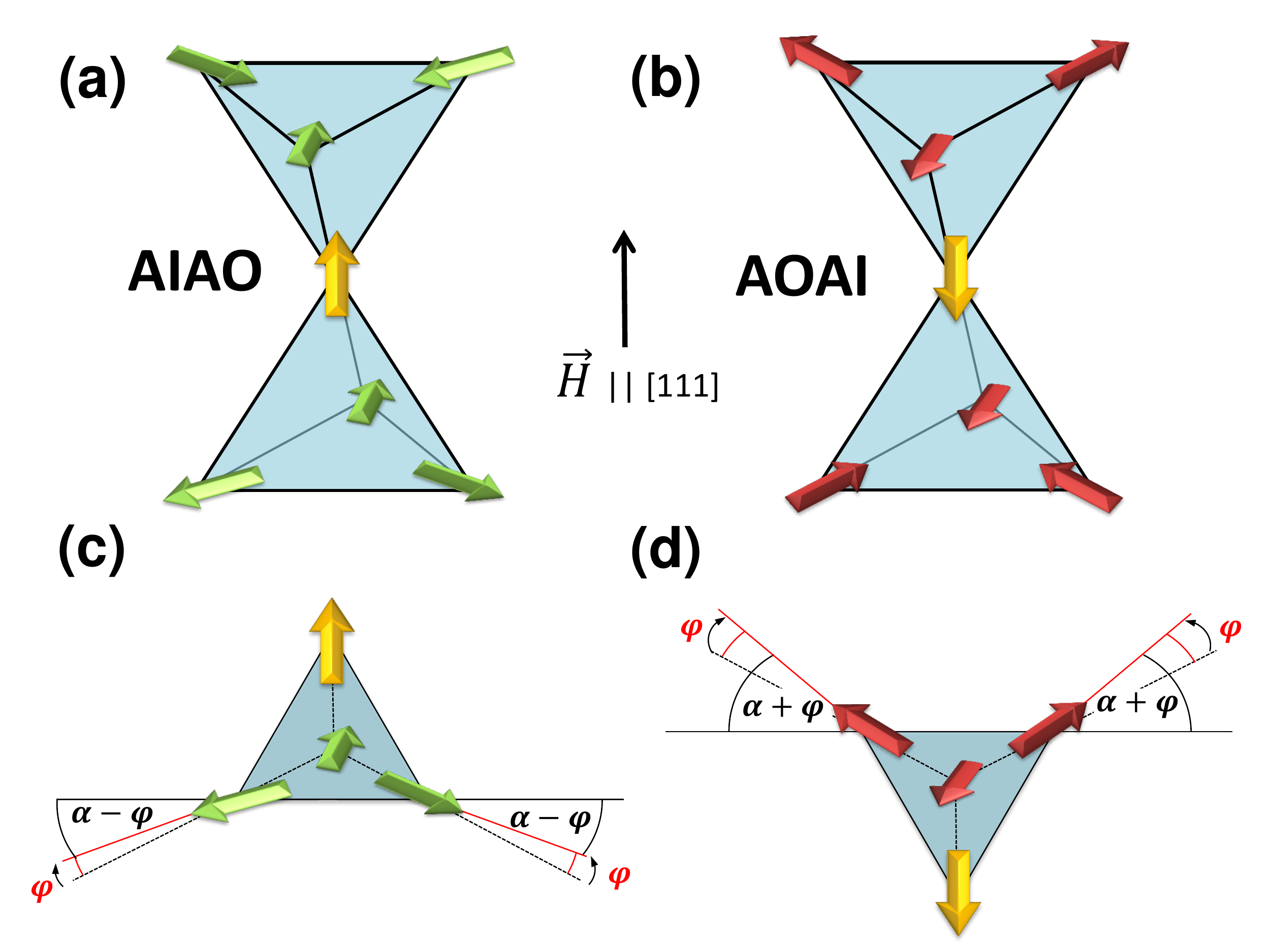}
	\caption{Visualization of the both possible configuration of the all-in-all-out ground state for an external magnetic field parallel to the [111] direction. In this case one quarter of the spins (yellow arrows) is aligned parallel (a) or anti-parallel (b) with respect to the field. The remaining spins have thus a component which is anti-parallel (green arrows) respectively parallel (red arrows) with respect to the field.This leads a canting of the non-collinear spins which is different for the AIAO (c) and AOAI domains (d).}
	\label{fig:AIAO-Canting_visualization}
\end{figure}

As mentioned above, the all-in-all-out state has two possible realizations. There are four distinct directions along which all spins are pointing. One of these local $\langle$111$\rangle$ directions is the [111] crystallographic direction so that all those spins are pointing either parallel (AIAO domain) or anti-parallel (AOAI domain) to an external magnetic field applied along [111] [Fig. \ref{fig:AIAO-Canting_visualization} (a),(b)].

Now we consider that in the presence of an external magnetic field, the spins can be canted by a small angle $\phi$ out of their local $\langle$111$\rangle$ anisotropy.
In the case of Nd$_2$Zr$_2$O$_7$, this assumption seems to be probable due to the admixture of non-Ising terms inside the CEF ground state of Nd$^{3+}$ (in addition to the leading term $\ket{^4I_{9/2},\pm 9/2}$) \cite{Xu_2015-1, Lhotel_2015}. 
It was further shown that the dipolar-octupolar nature of the Nd$^{3+}$ doublet leads both to the Ising-like $z$-component of the pseudo-spin and to a $y$-component in the local coordinate system of each spin \cite{Petit_2016-1}. Whereas the $z$-component is a result of the effective exchange interaction, $J_\mathrm{eff}$, the off-axis component is a result of the octupolar coupling $K$. By applying an external magnetic field, the off-axis component will be polarized. In this case, the pseudo-spin can be treated as a canted Ising-spin. One can further assume that the canting angle is equal for both domains because the canting strength is dominated by the ratio of $J_\mathrm{eff}$ and $K$.

Without such a spin canting, each pseudo spin $\vec{S}$ points along its local $\langle$111$\rangle$ direction.
Therefore, the total spin sum per tetrahedron $S_\mathrm{eff} = \sum_{i=1}^{4} \vec{S}_i \cdot \vec{\mathrm{e}_z}$ in the all-in-all-out state is zero as a necessary result of the antiferromagnetic order.
\begin{equation}
\begin{split}
S_\mathrm{eff} = S( +1 - 3\sin \alpha) = 0 , \qquad (\mathrm{AIAO})\\
S_\mathrm{eff} = S( -1 + 3\sin \alpha) = 0 . \qquad (\mathrm{AOAI})
\end{split}
\end{equation}
Whereas $\alpha$ is the angle between the perpendicular line of the [111] direction and the local $\langle$111$\rangle$ direction of the non-collinear spins and  $\sin{\alpha} = 1/3$. (For simplicity, the global $z$-direction was chosen to be parallel with respect to the [111] direction.)

\begin{figure}[t]
	\centering
	\includegraphics[width=\columnwidth]{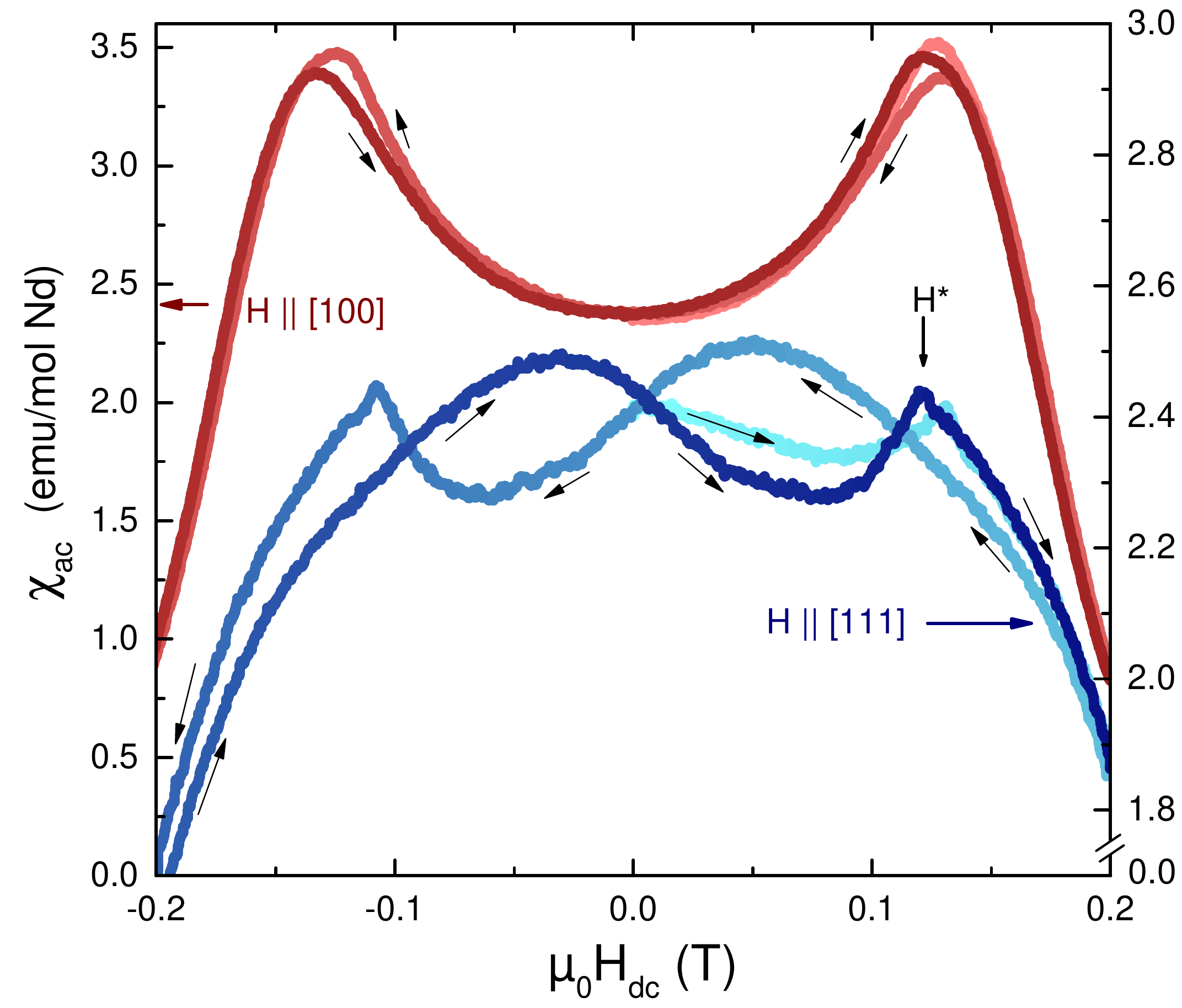}
	\caption{AC susceptibility versus external field. Magnetic hysteresis appears below the transition temperature for $H_{dc}$ $\|$ [111] (lower graph, $T$ = 70 mK) but for $H_{dc}$ $\|$ [100] (upper graph, $T$ = 107 mK) the curves almost coincide.}
	\label{fig:X_vs_B_compareOriantaions}
\end{figure}

A canting of the spins leads to a small ferrimagnetic contribution of each tetrahedron.
In the case for an AIAO domain, the three non-collinear spins have a component along the [111] crystallographic direction which is anti-parallel with respect to the field. For this reason, they cant towards the plane, perpendicular to the field [Fig. \ref{fig:AIAO-Canting_visualization} (c)]. In contrast, the three non-collinear spins of the AOAI domains have a parallel component with respect to the external field and cant towards a parallel configuration \cite{Arima_2012} [Fig. \ref{fig:AIAO-Canting_visualization} (d)].
If the assumed canting of the non-collinear spins in an fixed external magnetic field $B = \mu_0 H$ parallel to [111] has the fixed value $\phi$, the projection of total spin onto the field direction is given by
\begin{equation}
\begin{split}
S_\mathrm{eff} &= S( +1 - 3\sin (\alpha-\phi)) \qquad (\mathrm{AIAO})\\
&= S( +1 - \cos \phi + 2\sqrt{2} \sin \phi),\\
S_\mathrm{eff} &= S( -1 + 3\sin (\alpha+\phi)) \qquad (\mathrm{AOAI})\\
&= S( -1 + \cos \phi + 2\sqrt{2} \sin \phi).
\end{split}
\end{equation}
For the Zeemann energy per tetrahedron which is gained by the canting, follows $\Delta E = - \vec{m}\vec{B} = - g\mu_\mathrm{B}S_\mathrm{eff}B$.
One can than calculate the energy difference of the two all-in-all-out configurations per tetrahedron to be
\begin{equation}
\begin{split}
\Delta E &= E_\mathrm{AIAO} - E_\mathrm{AOAI} = -   2 g \cdot \mu_\mathrm{B} \cdot B \cdot (1 - \cos \phi)\\
&\approx - g\mu_\mathrm{B} B \cdot \phi^2 ,
\end{split}
\end{equation}
whereas the last approximation is valid for small canting angles $\phi \ll 1$, $\mu_\mathrm{eff}$ is the Bohr magneton and $g$ the effective g-factor of the Nd$^{3+}$ spins.
The twofold degeneracy of the all-in-all-out state is lifted in such a field and the AIAO arrangement is preferred.

The observed hysteresis, shown in Fig. \ref{fig:X_vs_B_Hpar111_all}, is, thus, a result of the changed domain structure. If the all-in-all-out state is realized in zero field, randomly distributed domains with the AIAO or AOAI configuration are formed. This was shown for example by use of microwave-impedance microscopy for Nd$_2$Ir$_2$O$_7$ \cite{Ma_2015} and X-ray diffraction for Cd$_2$Os$_2$O$_7$ \cite{Tardif_2015}. If an external magnetic field is applied parallel to the [111] direction, the spins are canted and the AIAO configuration becomes favorable. The system tends to transform the multi-domain structure to a single-domain phase.
The reduction of domain-wall density, where the antiferromagnetic order is broken, leads to a negative contribution of the magnetization and so to a decreasing susceptibility.
If the all-in-all-out structure is, on the other hand, formed under field cooled condition, a single AIAO domain appears.

If the external magnetic field is parallel to the [100] direction, almost no hysteresis can be seen for the field dependence of the AC susceptibility (see comparison for both orientations in Fig. \ref{fig:X_vs_B_compareOriantaions}). 
Whereas the spins are also canted in such a [100] field, the gained Zeeman energy is equal for both configurations of the all-in-all-out state because the AIAO configuration can be transformed to the AOAI configuration by a simple 90$^\circ$ rotation along the [100] axis \cite{Ma_2015}.
\begin{figure}[t]
	\centering
	\includegraphics[width=\columnwidth]{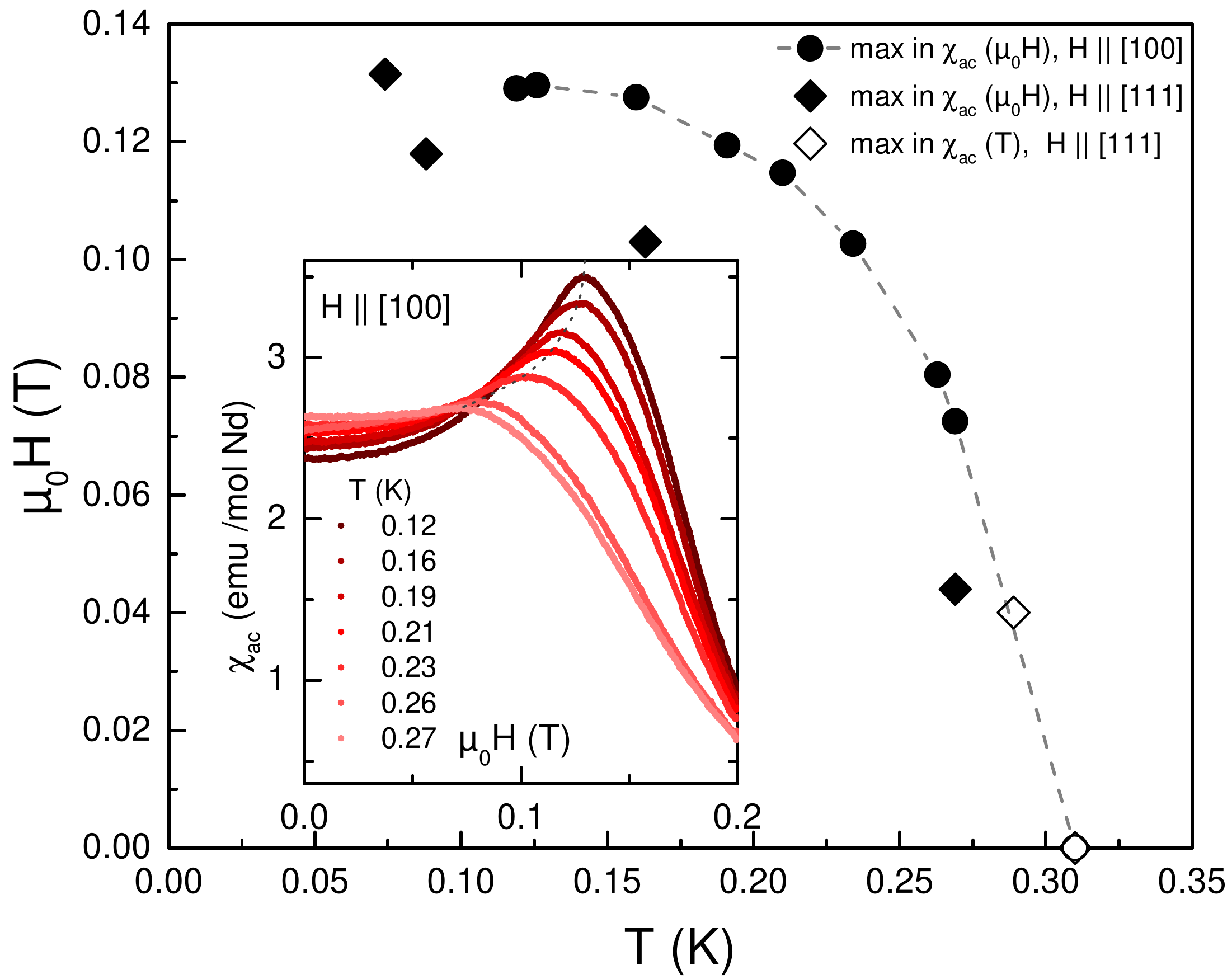}
	\caption{$H$-$T$ phase diagram of Nd$_2$Zr$_2$O$_7$. The critical fields and temperatures of the all-in-all-out state were obtained from the peak position of the temperature- and field-dependent AC susceptibilities for different orientations. The dashed line is a guide to the eye. Inset: AC susceptibility versus external static magnetic field  ($H$ $\|$ [100]) for different temperatures below the phase transition.}
	\label{fig:PhaseDiagram}
\end{figure}

From the maxima in the field- and temperature-dependent $\chi_\mathrm{ac}$ data, the phase diagram of Nd$_2$Zr$_2$O$_7$ can be constructed (Fig. \ref{fig:PhaseDiagram}).
The phase boundary is different for external DC fields applied along the [111] or [100] direction.
Whereas the critical field for suppressing the all-in-all-out order shows a steep increase near $T_N$, only a small further increase appears below 0.15 K. Such a shape of the all-in-all-out phase boundary was predicted by mean-field calculations considering the dipolar-octupolar nature of the Nd$_2$Zr$_2$O$_7$ doublet (see Fig. 5 in \cite{Lhotel_2015}) and is in global agreement with the phase diagram extracted from magnetization measurements \cite{Lhotel_2015}. 


\begin{figure}[t]
	\centering
	\includegraphics[width=0.7\linewidth]{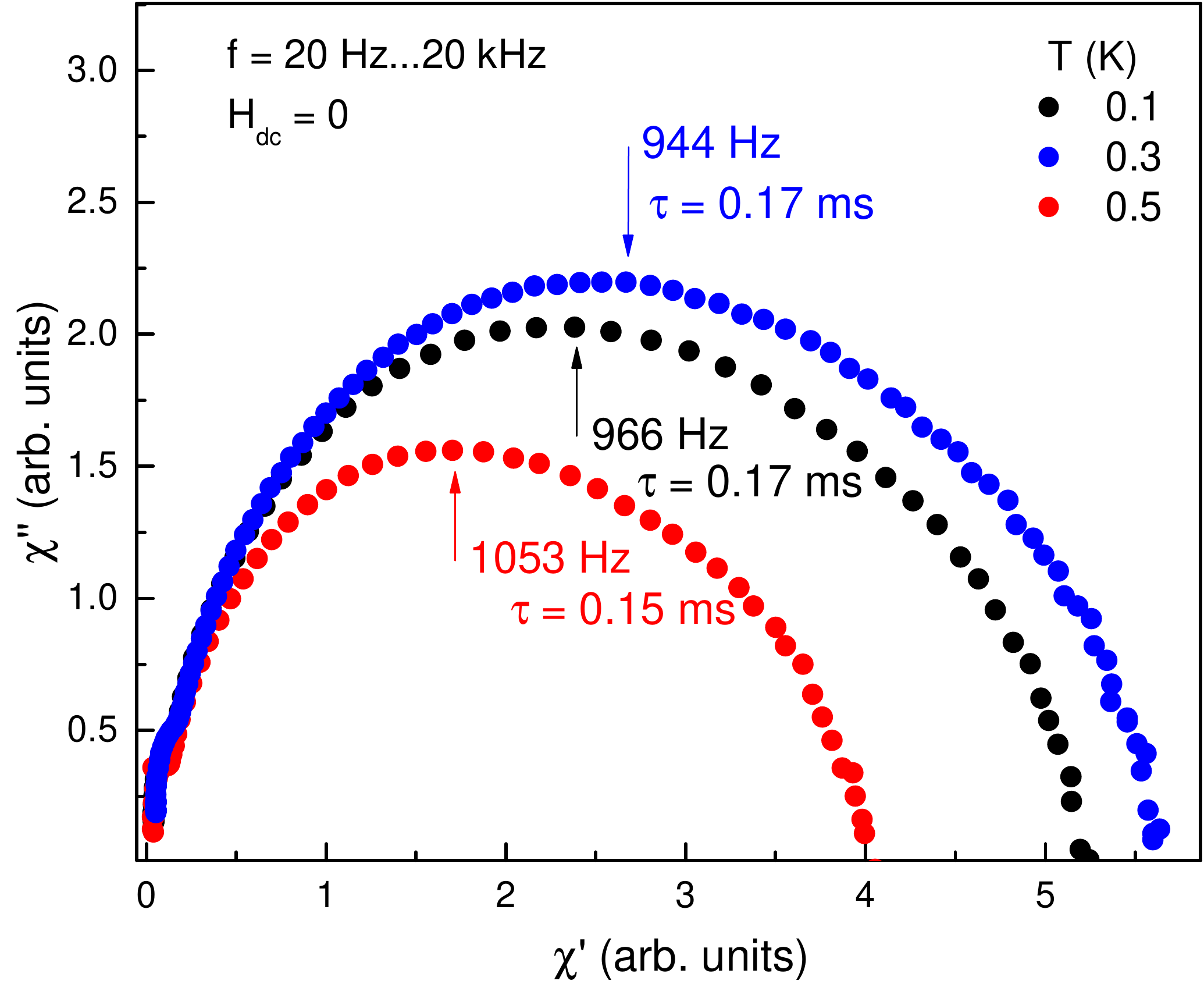}
	\caption{$\chi^{\prime\prime}$ versus $\chi^{\prime}$ at zero external magnetic field for several temperatures. The maximum in this Cole-Cole plot corresponds to $\omega\tau = 1$, where $\tau$ is the so-called single-spin relaxation time.}
	\label{fig:ACS_Cole-Cole}
\end{figure}
The spin-relaxation time for $H \|$ [111] is about $\tau$~=~0.2~ms and stays almost constant in the ordered and in the paramagnetic phase. This was determined by measuring the frequency dependence of $\chi_{ac}$ and analyzing the frequency of the maximum in the Cole-Cole plot, using $\tau \omega = 1$ with $\omega = 2\pi f$ (see Fig. \ref{fig:ACS_Cole-Cole}).
This value is fast in comparison to other pyrochlore systems, in particular compared to Dy$_2$Ti$_2$O$_7$ which has a much longer spin-relaxation time. At $T < 2$ K, $\tau$ of Dy$_2$Ti$_2$O$_7$ rapidly increases due to spin freezing \cite{Snyder_2004}.


In summary, Nd$_2$Zr$_2$O$_7$ undergoes a phase transition to an antiferromagnetically ordered all-in-all-out ground state configuration at temperatures below 0.31 K and magnetic fields smaller than 0.14 T. This result confirms recent calculations \cite{Onoda_2011, Lhotel_2015} as well as observations of neutron-diffraction experiments \cite{Xu_2015-1, Petit_2016-1}. 
The all-in-all-out state can be realized by two possible spin arrangements which leads to randomly distributed domains of both kinds in zero field. A canting of the spins out of their local $\langle$111$\rangle$ anisotropy leads to a gain of Zeeman energy. This energy is the same for the both configuration of the all-in-all-out state if the magnetic field is applied along the [100] direction, but different for a field along the [111] direction. The twofold degeneracy of the all-in-all-out state is, therefore, lifted for an applied field along [111] and the AIAO configuration is preferred.
The dynamics of the domain structure can be directly observed by measuring the resulting hysteresis in the field dependence of the AC susceptibility. To understand the spin canting mechanism in more detail, a study of a related Nd$_2T_2$O$_7$ compound with non-magnetic $T^{4+}$ ion, i.e. Nd$_2$Hf$_2$O$_7$, would be desirable. 

No signature for the recently proposed fragmentation of the magnetic moments in Nd$_2$Zr$_2$O$_7$ and the resulting spin-ice phase coexisting with the all-in-all-out ordering \cite{Petit_2016-1} was found. In contrast to spin-ice systems no frequency dependence of the dynamic susceptibility was found when changing $f$ from 50 Hz to 25 kHz. In addition, the spin relaxation time is fast compared to spin-ice systems such as Dy$_2$Ti$_2$O$_7$.

We believe that Nd$_2$Zr$_2$O$_7$ is a excellent model system to investigate and control domain structures of the antiferromagnetically ordered all-in-all-out state due to the small fields needed to change the domain structure and the fact that the magnetic moments are only located on the Nd$^{3+}$ sublattice.\\
\\
We acknowledge the Helmholtz Gemeinschaft for funding via the Helmholtz Virtual Institute (Project No. VH-VI-521) and DFG through Research Training Group GRK 1621 and SFB 1143. 
We also acknowledge support by HLD at HZDR, member of the European Magnetic Field Laboratory (EMFL).

\end{document}